\documentclass[12pt]{article}
\usepackage{amsmath}
\usepackage{amsfonts}
\usepackage{amssymb}
\newtheorem{theorem}{Theorem}

\newtheorem{proposition}[theorem]{Proposition}

\newenvironment{proof}[1][Proof]{\textbf{#1.} }{\ \rule{0.5em}{0.5em}}

\begin{document}

\title{PROGRAMS WITH STRINGENT PERFORMANCE OBJECTIVES WILL OFTEN EXHIBIT CHAOTIC BEHAVIOR}
\author{M. CHAVES\\Escuela de Fisica, Universidad de Costa Rica\\San Jose, Costa Rica\\mchaves@cariari.ucr.ac.cr}
\date{March 15, 1999}
\maketitle
\begin{abstract}
Software for the resolution of certain kind of problems, those that rate high
in the \emph{Stringent Performance Objectives} adjustment factor (IFPUG
scheme), can be described using a combination of game theory and autonomous
systems. From this description it can be shown that some of those problems
exhibit chaotic behavior, an important fact in understanding the functioning
of the related software. As a relatively simple example, it is shown that
chess exhibits chaotic behavior in its configuration space. This implies that
static evaluators in chess programs have intrinsic limitations.
\end{abstract}

\section{Introduction}

\noindent
%
%
IBM's Deep Blue, a powerful chess playing machine consisting of two
parallel-process tandem supercomputers programmed by a team of experts lead by
team manager C. Tan [Hsu \emph{et al.}, 1990; Horgan, 1996; Hsu, 1990; Slate,
1984], played the world chess champion G. Kasparov several games in 1996 and
1997 with fairly even results. Actually, programmer Hsu's estimate back in
1990 of the future machine's playing strength was 4000 ELO points (chess'
rating system), far greater than Kasparov's $\sim$2800 present rating. In
three minutes, which is the game's average pondering time, the machine could
calculate 20 billion moves, enough to for a 24-ply search and an up to 60-ply
search in critical tactical lines. Since grandmasters can calculate just a few
moves ahead, it seems very peculiar that a human could hold his own on the
face of such an overwhelming opposition.

In this paper we are interested in a special kind of problem and the software
written for it. It is the kind of problem whose software would score high in
the \emph{Stringent performance objectives} [Abran \& Robillard, 1996]
adjustment factor of the International Function Point User's Group (IFPUG).
Examples are, for instance, the control of air-traffic at a busy airport, the
scheduling of trains in areas with heavy traffic, and field military
operations. One way of approaching this kind of problem is to treat it within
the context of game theory, as a 2-player game. The first player would be the
comptroller, central authority or headquarters, and the second is the system
itself, that acts and reacts out of its own nature. The first player pursues
to maintain control of a complicated system by choosing its moves, that is, by
affecting the system in the ways available to him. He would like the system to
always remain in states such that certain state variables (they could be
safety, efficiency, lethality or others) are kept extremized. The performance
objectives would be to extremize these state variables.

The nature of this kind of problem is such that it is necessary to see ahead
what is going to happen. At least in theory, the first player must have
arrived at his move only after having taken into consideration all of the
possible responses of the second player. This is a common situation in game
theory, and is another reason why the language of game theory is very
well-suited to discuss both this kind of problem and the software developed to
help deal with it. The typical program contains two fundamental sectors:

\begin{enumerate}
\item \emph{a ply calculator,} that is able to look ahead at all possible
continuations of the tree a certain number of plies ahead,

\item \emph{a static evaluator,} that gives an evaluation of the resulting
state of the problem at the end of each branch of plies.
\end{enumerate}

\-Although there are many different ways of programming the ply calculator or
the static evaluator, their complementary, basic functions are clear: the
first is a brute force calculator of all possible states to come, and the
second is an evaluator of the final resulting state of the system,
intrinsically and on the basis of the state itself, without any resort to
further calculation.

The different states of the problem can be seen as a function of time. If one
is able to express each state using a mathematical description, the problem of
the time-development of the state while maintaining certain state variables
extremized can be described as an autonomous system. If the equations
describing the time-development of the system are nonlinear, it is very likely
that the problem is going to exhibit chaotic [Alligood \emph{et al.}, 1997]
behavior. Therefore, the software for these problems has an intrinsic
limitation on its accuracy, even if it still may be extremely useful.

As an example we will work out the case of chess, a useful one because, while
nontrivial, it is not nearly as complex as some of the other problems are.
Chess' software scores high in the \emph{Stringent performance objectives}
adjustment factor. We will prove that chess exhibits chaotic behavior in its
configuration space and that this implies its static evaluators possess
intrinsic limitations: there are always going to be states or positions that
they will not be able to evaluate correctly. It is likely that this is
precisely the explanation for the peculiar situation mentioned in the first
paragraph: that a human being can hold his own at a chess game with a
supercomputer. The ply calculator part of the program of the supercomputer
would be tremendously effective, but the static evaluator would not be so faultless.

\section{An abstract mathematical representation of chess}

We have to describe each possible state (or position) in chess. To describe a
particular state we shall use a 64 dimensional vector space, so that to each
square of the board we associate a coordinate that takes a different value for
each piece occupying it. A possible convention is the following:

\begin{itemize}
\item  A value of zero for the coordinate of the dimension corresponding to a
square means that there is no piece there.

\item  For the White pieces the convention would be: a value of 1 for the
coordinate means the piece is a pawn, of 2 means it is a pawn without the
right to the \emph{en passant move,} of 3 that it is a knight, of 4 a bishop,
of 5 a rook, of 6 a queen, of 7 a king, and of 8 a king without the right to castle.

\item  The values for the Black pieces would be the same but negative.
\end{itemize}

Let us represent the 64-component vector by the symbol $x$. A vector filled
with the appropriate numbers can then be used to represent a particular state
of the game. We shall call the 64-dimensional space consisting of all the
coordinates \emph{the configuration space C} of the game. The succeeding moves
of a pure strategy can be plotted in this space, resulting in a sequence of
points forming a path.

Now we construct a function $f:C\rightarrow C$ that gives, for any arbitrary
initial state of a chess game, the control strategy to be followed by both
players. The existence of this function is assured by the Zermelo-von Neumann
theorem [von Neumann \& Morgenstern, 1944] that asserts that a finite 2-person
zero-sum game of perfect information is strictly determined, or, in other
words, that a pure strategy exists for it. For a given initial chess state
this means that either

\begin{itemize}
\item  White has a pure strategy that wins,

\item  Black has a pure strategy that wins,

\item  both are in possession of pure strategies that lead to a forced draw.
\end{itemize}

Consider a certain given initial state of the game where White has a pure
strategy leading to a win. (The two other cases, where Black has the win or
both have drawing strategies can be dealt with similarly and we will not treat
them explicitly.) Let the initial chess state be given by the 64-component
vector $x_{0}$, where we are assuming that White is winning. The states
following the initial one will be denoted by $x_{n}$, where the index is the
number of plies that have been played from the initial position. Thus $x_{1}$
is the position resulting from White's first move, $x_{2}$ is the position
resulting from Black's first move, $x_{3}$ is the position resulting from
White's second move, and so on. Since White has a winning pure strategy, it is
obvious that, given a certain state $x_{n}$, $n$ even, there must exist a
vector function $f$ so that, if $x_{n+1}$ is the next state resulting from
White's winning strategy, then $f(x_{n})=x_{n+1}$. On the other hand, if $n$
is odd, so that it is Black's turn, then we define $f$ to be that strategy for
Black that makes the game last the longest before the checkmate. Again, the
pure strategy that is available to Black according to the Zermelo-von Neumann
theorem allows us to define a function $f(x_{n})=x_{n+1}$. The function $f$ is
thus now defined for states with $n$ both even and odd.

The function $f$ allows us to define another function $g:C\rightarrow C$,
\emph{the control strategy vector function} [Abramson, 1989], defined by
$g(x_{n})=f(x_{n})-x_{n}$. With it we can express the numerical difference
between the vectors corresponding to two consecutive moves as follows:%
\begin{equation}
g(x_{n})=x_{n+1}-x_{n}. \tag{1}%
\end{equation}
Given any initial state $x_{0}$, this function gives us an explicit control
strategy for the game from that point on.

\section{Chaos in configuration space}

A set $N$ of simultaneous differential equations,%
\begin{equation}
g(x)=\frac{dx}{dt}, \tag{2}%
\end{equation}
where $t$ is the (independent) time variable, $x\in R^{N}$ and the $g$ are
known $g:R^{N}\rightarrow R^{N}$ vector functions, is called an autonomous
system [Alligood \emph{et al.}, 1997]. The time $t$ takes values in the
interval $0\leq t\leq T$. Let us discretize this variable, as is often done
for computational purposes [Parker \& Chua, 1989]. We assume it takes only
discrete values $t=0,\Delta t,2\Delta t,\ldots,T$. After an appropriate
scaling of the system one can take the time steps to be precisely $\Delta
t=1$. Let the initial condition of the system be $x(0)\equiv x_{0}$, and let
us define $x(1)\equiv x_{1},$ $x(2)\equiv x_{2}$, and so on. By taking $N=64$
one can then rewrite (2) in a form that is identical to (1).

Nonlinear autonomous systems in several dimensions are always chaotic, as
experience shows. Is the control strategy function nonlinear? A moment's
consideration of the rules of the game tell us that the control function has
to be nonlinear and that, therefore, the system described by (1) has to be chaotic.

For some kinds of chess moves the difference $x_{n+1}-x_{n}$ has a relatively
large value that would correspond to a jerky motion of the system, and the
question can be raised if such a motion could really occur in a an autonomous
system. But the important thing to realize is that if \emph{even} typical
autonomous nonlinear systems (that possess a smooth function $g(x)$) do show
chaotic behavior, then \emph{certainly} the system that represents chess (with
a jerky control strategy function) should also show it.

The chaotic nature of the paths in configuration space has several immediate
implications, but certainly one of the most interesting is the following:

\begin{proposition}
It is not possible to program a static evaluator for chess that works
satisfactory on all positions.
\end{proposition}

\begin{proof}
The point of the proposition is that a program with a good static evaluator is
always going to have shortcomings: it will always evaluate incorrectly at
least some positions. If one programs another static evaluator that evaluates
correctly these positions, one will notice soon that there are others that the
new program still cannot evaluate correctly. In last analysis the perfect
evaluator for chess would have to be an extremely long program, and for more
complex systems of this kind, an infinite one. To see it is not possible to
program a static evaluator for chess that works correctly on all positions,
notice that it would have to evaluate on the basis of the state itself
\emph{without recourse to the tree.} The evaluation of the state has to be
done using heuristics, that is, using rules that say how good a state is on
the basis of the positions of the pieces and not calculating the tree. But
this is not possible if chess is chaotic because then we know the smallest
difference between two states leads to completely diverging paths in
configuration space, that is, to wholly differing states a few plies later.
Therefore the heuristic rules of the static evaluator have to take into
account the smallest differences between states, and the evaluators have to be
long or infinite routines. Static evaluators, on the other hand, should be
short programs, since they have to evaluate the states at the end of each
branch of the tree.
\end{proof}

Another interesting point is that chaos exacerbates the horizon effect
[Berliner, 1973]. This is the problem that occurs in game programming when the
computer quits the search in the middle of a critical tactical situation and
thus it is likely that the heuristics return an incorrect evaluation [Shannon,
1950]. In a sense, what the proposition is saying is that practically all
states are critical, and that the horizon effect is happening all the time and
not only for some supposedly very special positions.

\section{Comments}

We have seen that it is likely that the pure strategy paths of chess in
configuration space follow chaotic paths. This implies that practical static
evaluators must always evaluate incorrectly some of the states. As a result
the horizon problem is exacerbated.

The reason why a machine such as Deep Blue is not far, far stronger than a
human has to do again with the problem of programming a static evaluator. Even
though the machine searches many more plies than the human does, at the end of
each branch it has to use a static evaluator that is bound to incorrectly
evaluate some states. This adds an element of chance to the calculation. The
fact that Deep Blue at present has a playing strength similar to the best
human players tells us that the human mind has a far better static evaluator
than Deep Blue (assuming one can apply these terms to the human mind). If
chess were not chaotic the overwhelming advantage in ply calculation that the
machine has would allow it to play much better than any human could.

In practice, of course, as long as computers keep getting faster and having
more memory available, it is always possible to keep improving the static
evaluators. If computers can be programmed to learn from their experience they
could improve their static evaluators themselves. This was the idea of the
program for another game, link-five [Zhou, 1993].

Now, in a general vein, it should be clear why programs that would score high
in the IFPUG's \emph{Stringent performance objectives} adjustment factor would
tend to be exhibit chaotic behavior in their configuration spaces. The
software of this type of program has to foresee what is going to happen while
extremizing certain state variables, as we mentioned before. This kind of
problem is equivalent to an autonomous system of differential equations that
exhibits chaos, so that the control strategy vector function $g$ of the system is extremely sensitive to the smallest differences in a state $x$. Thus any static evaluator that one programs (that has to be heuristic in nature) is going to be severely limited.

Nevertheless, the consideration we made two paragraphs ago for chess is also
true for this kind of problem in general: as long as computers get faster and
have more memory the programs can be prepared to deal with more and more
situations. Rules of thumb that humans have learned from experience can be
added to the evaluators. Alternatively, the programs can be written so that
they learn from their experience. But they are always going to be \emph{very} long programs.

\bigskip

\bigskip

\noindent\textbf{References}

\bigskip

\noindent Abramson, B. [1989] ``Control strategies for two-player games'', ACM
Comp. Surveys \textbf{21}, 137- 161.

\bigskip

\noindent Abran, A. \& Robillard, P. N. [1996] IEEE Trans. Software Eng.
\textbf{22}, 895-909.

\bigskip

\noindent Alligood, K. T., Sauer, T. D. \& Yorke J. A. [1997] \emph{Chaos: An
Introduction to Dynamical Systems} (New York, Springer-Verlag).

\bigskip

\noindent Berliner, H. J. [1973] ``Some necessary condition for a master chess
program'', \emph{Proceedings of the 3rd International Joint Conference on
Artificial Intelligence, Stanford, CA} (Los Altos, Morgan Kaufmann), 77-85.

\bigskip

\noindent Horgan, J. [1996 ] ``Plotting the next move'', Scien. Am.
\textbf{274} no. 5, 10.

\bigskip

\noindent Hsu, F. [1990] ``Large scales parallelization of alpha-beta search:
an algorithmic and architectural study with computer chess'', Ph.D. thesis.
Carnegie-Mellon University Computer Science Department, CMU-CS-90-108.

\bigskip

\noindent Hsu, F., Anantharaman, T., Campbell M., \& Nowatzyk, A. [1990] ``A
Grandmaster Chess Machine'', Scient. Am. \textbf{263} no. 4, 44-50.

\bigskip

\noindent von Neumann, J. \& Morgenstern O. [1944] \emph{Theory of Games and
Economic Behavior} (Princeton, Princeton University Press).

\bigskip

\noindent Parker, T.S. \& Chua L.O. [1989] \emph{Practical Numerical
Algorithms for Chaotic Systems} (New York, Springer-Verlag).

\bigskip

\noindent Shannon, C.E. [1950] ``Programming a computer for playing chess'',
Philos. Mag. \textbf{41}, 256-275.

\bigskip

\noindent Slate, D. J. \& Atkin, L. R. [1984] ``Chess 4.5-The Northwestern
University chess program'', in $\emph{Chess}$\emph{ Skill in Men and Machine},
ed. Frey, P. W. (New York, Springer Verlag).

\bigskip

\noindent Zhou, Q. [1993] ``A distributive model of playing the game link-five
and its computer implementation'', IEEE Trans. Syst., Man, Cybern.,
SMC-\textbf{23}, 897-900.
\end{document}